\documentclass{article}
\usepackage{spconf,amsmath,graphicx}
\usepackage{hyperref}
\usepackage{color}
\usepackage{xcolor}
\usepackage{url}
\usepackage{subcaption}
\usepackage{algorithm}
\usepackage{algorithmic}
\usepackage{amsthm}
\usepackage{graphicx}
\usepackage{tabularx}
\usepackage{amssymb}
\usepackage{booktabs}
\usepackage{amsmath}
\usepackage{wrapfig}
\usepackage{multirow}
\usepackage{multicol}
\usepackage{color, colortbl}
\usepackage[normalem]{ulem}
\usepackage{enumitem}
\setlist{nosep, leftmargin=14pt}

\usepackage{mwe} 

\definecolor{blue_p}{rgb}{0.185, 0.43, 0.73}
\definecolor{orange_p}{rgb}{0.92, 0.61, 0.34}
\definecolor{aliceblue}{rgb}{0.94, 0.97, 1.0}
\definecolor{airforceblue}{rgb}{0.43, 0.49, 0.7}
\definecolor{babyblueeyes}{rgb}{0.83, 0.89, 1}
\newlength{\bibitemsep}
\newlength{\bibparskip}
\let\oldthebibliography\thebibliography
\renewcommand\thebibliography[1]{%
  \oldthebibliography{#1}%
  \setlength{\parskip}{\bibitemsep}%
  \setlength{\itemsep}{\bibparskip}%
}
\title{Unsupervised Airway Tree Clustering with Deep Learning: The Multi-Ethnic Study of Atherosclerosis (MESA) Lung Study}

\name{Sneha N. Naik$^{1}$ , Elsa D. Angelini$^{1,3}$, R. Graham Barr $^{1}$, Norrina Allen$^{4}$, Alain Bertoni $^{5}$, Eric A. Hoffman$^{6}$ , Ani Manichaikul $^{7}$, Jim Pankow$^{8}$, Wendy Post$^{9}$, Yifei Sun$^{1}$, Karol Watson$^{10}$, Benjamin M. Smith$^{2}$, Andrew F. Laine$^{1}$}
\address{$^{1}$ Columbia University, USA. $^{2}$ McGill University, Canada.\\$^{3}$ Telecom Paris LTCI, France. $^{4}$ Northwestern University, USA. $^{5}$  Wake Forest University, USA.\\
$^{6}$ University of Iowa, USA. $^{7}$ University of Virginia, USA.
$^{8}$ University of Minnesota, USA. \\$^{9}$ Johns Hopkins University, USA.
$^{10}$ University of California, USA.
}
\begin{document}

\maketitle

\begin{abstract}
High-resolution full lung CT scans now enable the detailed segmentation of airway trees up to the 6th branching generation. The airway binary masks display very complex tree structures that may encode biological information relevant to disease risk and yet remain challenging to exploit via traditional methods such as meshing or skeletonization. Recent clinical studies suggest that some variations in shape patterns and caliber of the human airway tree are highly associated with adverse health outcomes, including all-cause mortality and incident COPD. However, quantitative characterization of variations observed on CT segmented airway tree remain incomplete, as does our understanding of the clinical and developmental implications of such. In this work, we present an unsupervised deep-learning pipeline for feature extraction and clustering of human airway trees, learned directly from projections of 3D airway segmentations. We identify four reproducible and clinically distinct airway sub-types in the MESA Lung CT cohort. 
\end{abstract}
\section{Introduction}
\label{sec:intro}
A growing body of evidence suggests that differences in airway tree structure are associated with adverse health outcomes \cite{vameghestahbanati2023association}\nocite{smith2018human, smith2020association, bodduluri2021computed}-\cite{bodduluri2018airway}. Airway tree caliber assessed on CT is a risk factor for all-cause mortality, in addition to cause-specific mortality from COPD, lung cancer and even atherosclerotic cardiovascular disease \cite{vameghestahbanati2023association}. Variations in airway structure are evident at the segmental level in over 25\% of the population \cite{smith2018human}. The presence of accessory segmental airway or absence of standard segmental airway is associated with higher odds of COPD, and the latter is also associated with genetic marker FGF10, which regulates embryonic airway budding \cite{smith2018human}. Airway structure further impacts the deposition of inhaled particles at bifurcations \cite{lambert2011regional}. All of the above motivate the identification of airway tree subtypes on large populations of segmented CT scans and the analysis of their clinical relevance.\\
Prior work analyzed airway topology in Billera-Holmes-Vogtmann (BHV) tree space \cite{billera2001geometry} for clustering \cite{wysoczanski2021unsupervised}, and classification of COPD subjects based on geodesic distance between tree topologies \cite{feragen2013tree}. 

However, metrics in BHV tree-space require that the `leaf set' be conserved across all trees in the dataset, thereby limiting the metric to the depth of anatomical labeling. BHV tree-space further considers only topology and not the impact of airway caliber. 
Improvements in airway segmentation from CT have accelerated in recent years \cite{nadeem2020ct, zhang2023multi} and robust airway segmentations can now be obtained beyond the segmental level. This leads to large-size segmentation masks with fine details, which need to be analyzed on both lungs together (rather than on patches) to compare subjects. \\ 

In this work, we propose to use the deep-learning (DL) auto-encoding capacity to learn airway tree structural patterns directly from projections of whole 3D airway segmentation masks.
Our method is designed to handle the following challenges: (1) Potential detrimental influence of the trachea which comprises $\approx$40\% of voxels in a 3D airway mask. This might lead to disproportional importance of this structure in DL encoder training while at the same time there is between-study variability of the superior cut-off plane position in the CT field of view. 
(2) Lack of co-registration between subjects CT scans. 
(3) Tradeoff required between voxel-level segmentation (usually done on patches) and encoding of features at scan level for population clustering and proposition of new airway shape phenotypes. \\ 
We train and evaluate our proposed method on binary airway masks observed on HRCT scans from the Multi-Ethnic Study of Atherosclerosis (MESA) \cite{bild2002multi} Exam 5 participants to identify airway shape phenotypes. 

\section{Materials and Methods}
\label{sec:materials_methods}

\subsection{Datasets and airway segmentations}\label{subsec:dataset_mesa}
We exploited two CT scan cohorts with airway segmentations:
(1) The \textbf{MESA Lung study} \cite{rodriguez2010association} is part of a population-based study of 6,814 participants recruited in 2000-2002 across six study centers in the USA. Participants were ages 48-85, were 53\% female, 39\% white, and were free of cardiovascular disease at baseline \cite{bild2002multi}. In 2010-2012, the MESA Lung Study performed full-lung high-resolution CT scans for 3,195 participants with in-plane resolution in range $\left[0.4668, 0.9180\right]$mm, and slice spacing $0.5$mm following the SPIROMIC CT protocol \cite{sieren2016spiromics}. 
Airway segmentation masks (from VIDA Diagnostics) are available with on average $323 \pm 124$ airway segments per tree, each with a maximum depth of $12.7 \pm 1.6$ Weibel generations (0 = trachea, 16 = terminal bronchioles). 
(2) The \textbf{ATM'22} grand challenge dataset consists of 500 anonymized CT scans, of which n=300 have reference airway segmentations \cite{zhang2023multi}, showing on average $202\pm73$ segments per tree with a maximum depth of $11.1 \pm 1.6$. Lung conditions of the scanned subjects range from healthy status to severe pulmonary disease.

\subsection{Airway masks pre-processing}\label{subsec:preproc}

From the MESA Lung cohort, we retained n=2,587 participants based on the criteria: (1) the airway mask contains a single connected component, with airways at least to the segmental level; (2) Each lobe in the lung mask comprises $\geq 5\%$ of the total lung volume; (3) Segmental airway branch variant annotations from \cite{smith2018human} are available. 
We rotated the full-size 3D airway masks to align their first three PCA vectors with the image axis and then cropped with the bounding box around the segmentation. We then computed Maximum intensity projections (MIPs) for the axial, coronal, and sagittal planes. 
\begin{figure}[htb]
 \resizebox{\linewidth}{!}{
\begin{minipage}[b]{1.0\linewidth}
  \centering
  \centerline{\includegraphics[width=5cm]{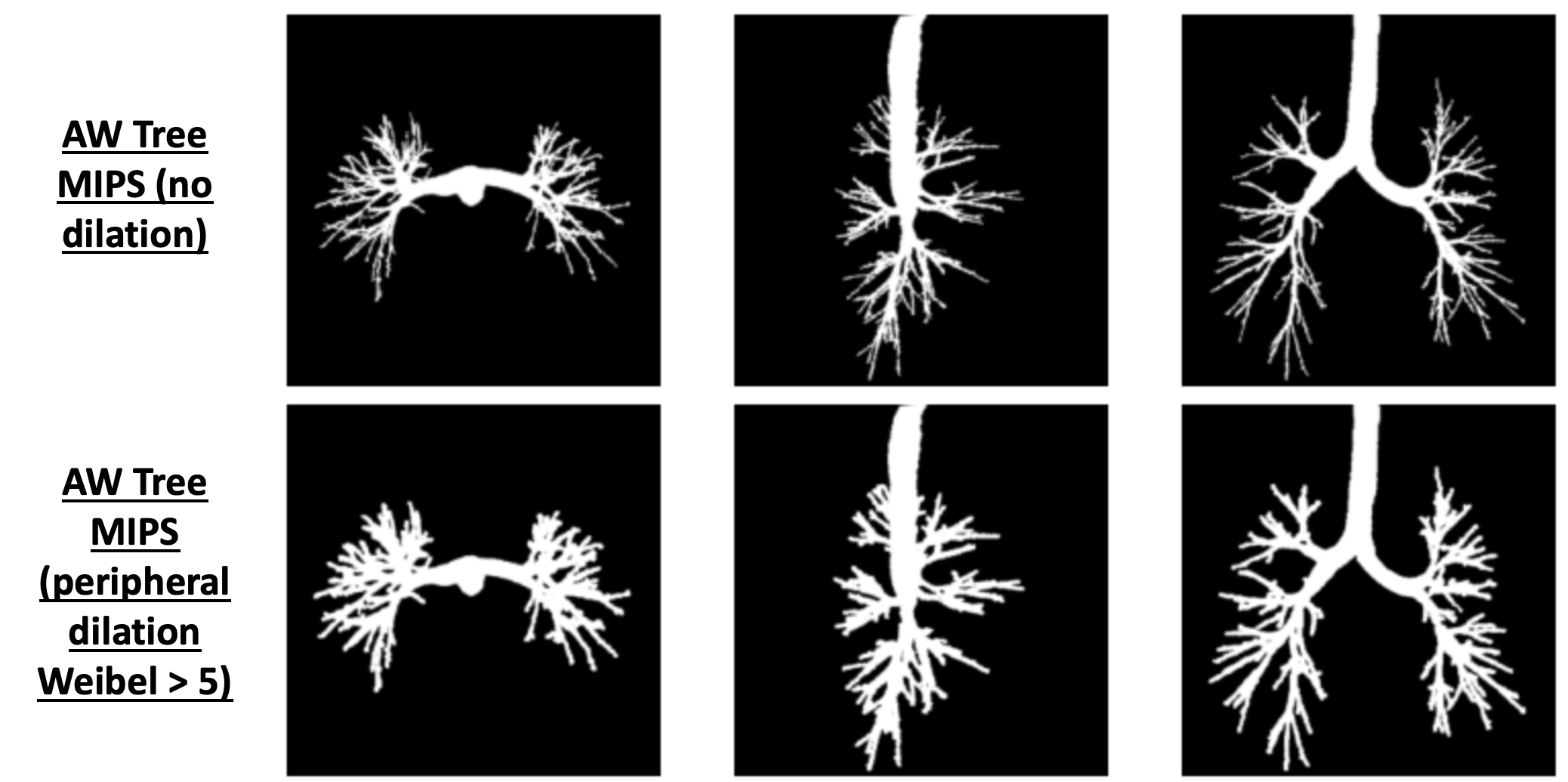}}

\end{minipage}}
\caption{2D projections of airway masks in axial, coronal, and sagittal planes without and with peripheral airways dilations to emphasize shape characteristics of small airways.}
\label{fig:dil_aw_tree}
\end{figure}
We tested a scenario where we increased the visibility and hence up-weight the importance of peripheral airways in the MIP representations. We generated 2 separate MIPs from airways above or below Weibel generation $>5$. For the generation above 5, we applied a dilation kernel (4x4 ones) and then merged it with the other MIP. We downsampled each MIP and concatenated the 3 views to an array of size 3 x 256 x 256 per subject. We generated MIPs with (T) and without (NT) the trachea and with dilation (D) or without (ND).

\subsection{Deep encoding of airway MIPs}\label{subsec:DL_model}
\begin{figure}[htb]
 \resizebox{\linewidth}{!}{
\begin{minipage}[b]{1.0\linewidth}
  \centering
  \centerline{\includegraphics[width=8.5cm]{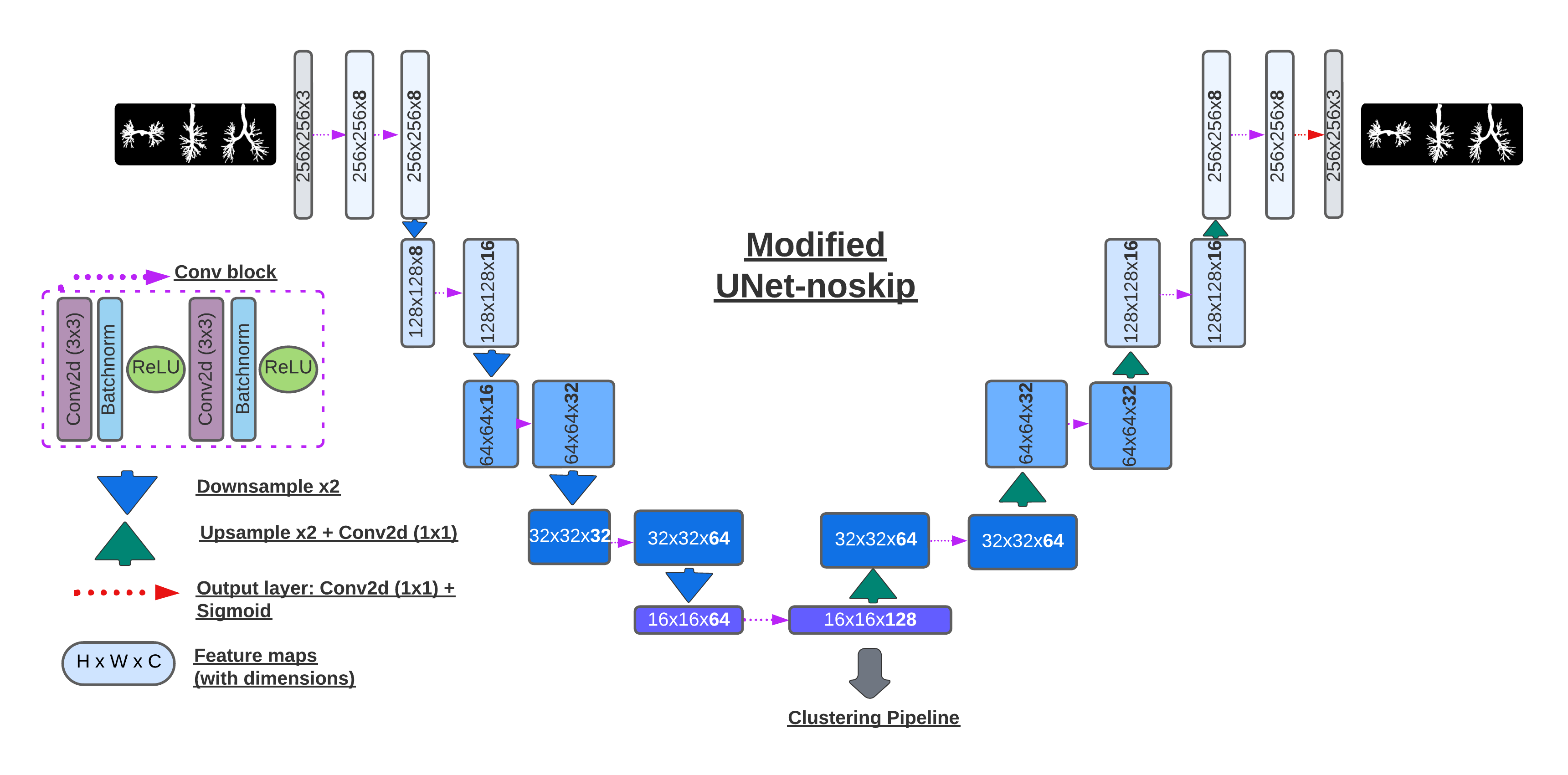}}

\end{minipage}}
\caption{Modified UNet architecture (UNet-noskip). }
\label{fig:unet}
\end{figure}
For the auto-encoder, we modified the baseline UNet architecture \cite{ronneberger2015u} by: (1) reducing the feature maps per layer by $1/8$th; (2) eliminating skip connections, and constraining all reconstruction information to a bottleneck feature of size 128x16x16, which we used for downstream clustering. The UNet-noskip model has 537k tunable parameters (Figure \ref{fig:unet}). 
We split the 2,587 subjects into 5 train-validation folds, stratified by segmental airway variant \cite{smith2018human}. 
The reconstruction loss function was an equally weighted combination of Binary Cross Entropy and Dice Losses. For each fold, we conducted mini-batch gradient descent with a batch size of 12, Adam optimizer, a learning rate of 0.001, and a cosine annealing learning rate scheduler with warm restarts. The learning rate decayed from 0.001 to 0 over 20 epochs, restarted, and subsequently decayed over 40 epochs. The decay period roughly doubled at each restart. Early stopping of training was conducted if there was no improvement in validation loss after 10 epochs. 
We selected the UNet-noskip model from the highest-performing fold for fine-tuning for 5 epochs.

We reduced the learning rate to 0.0001, with all other parameters as above. 

We used the ATM grand challenge dataset for external evaluation of the trained autoencoder.

\subsection{Airway tree clustering}\label{subsec:cluster_method}

We applied PCA to reduce the UNet-noskip bottleneck feature size from 128x16x16 per tree to 2,048 (98\% variance explained); We computed pairwise L2 distances between vectors to generate a $kNN$ graph with edge weights between two airway trees $E(i,j) = \alpha(i,j) \sqrt{|v_i - v_j|^2}$, where $\alpha(i,j) = 1$ if tree $j$ is one of the $k$ nearest neighbors of tree $i$ and otherwise 0. We used the Louvain algorithm \cite{de2011generalized}  to perform community detection on the $kNN$ graph by iteratively selecting partitions that optimize a `modularity' score on the graph. Increasing $k$ in the pipeline generally results in fewer clusters. We set the value of $k$ empirically by running Louvain clustering with $k \in [5,2500]$ and selecting the first plateau where the number of clusters remains constant for $\Delta k = 100$. We set $k_{opt}$ as the midpoint of the plateau (Figure \ref{fig:tune_kopt_reprod}a). 

We also report clustering reproducibility with the following variants: (1) clustering on 80\% rather than 100\% of training data; (2) varying $k$ in $k$NN graph; (3) using Cosine distance in $k$NN graph (4) clustering on PCA-based 1024-D vectors (89\% variance explained).

Finally, we related airway tree clusters to spirometry-based measures of chronic obstructive pulmonary disease (COPD) and CT-based measures of emphysema severity and dysanapsis by computing the least square mean values adjusted for participant age, sex, height, race-ethnicity, study site, smoking status, and pack-years smoking.

\section{Results}

\begin{figure*}[htb]

\resizebox{\linewidth}{!}{
\begin{minipage}[b]{.49\linewidth}
  \centering
    \resizebox{\linewidth}{!}{
      \centerline{\includegraphics[width=8.5cm]{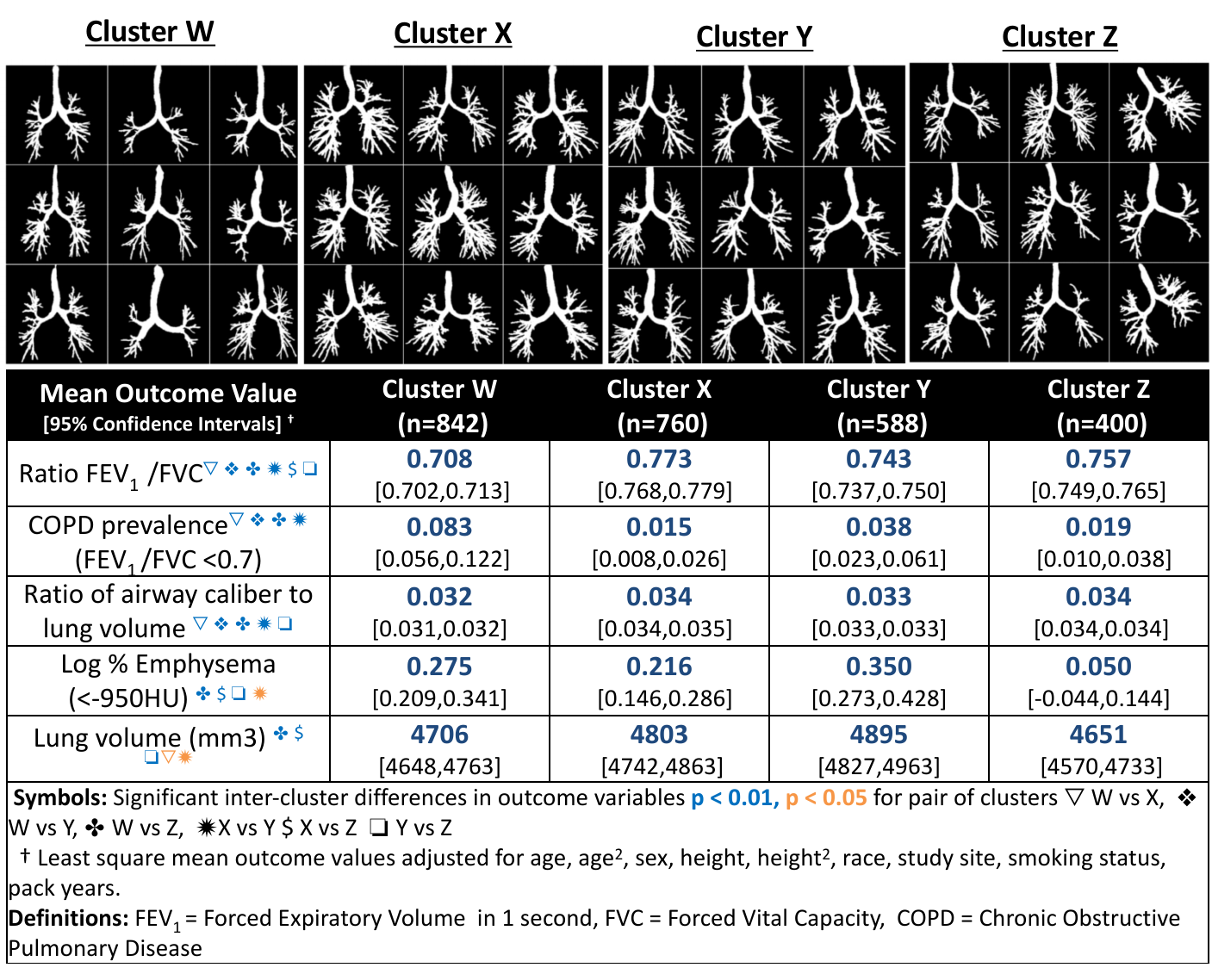}}}

  \centerline{(a) Random samples per cluster from $\mathcal{C}(D+T)$}\medskip
\end{minipage}
\hfill
\begin{minipage}[b]{0.49\linewidth}
  \centering
    \resizebox{\linewidth}{!}{
  \centerline{\includegraphics[width=8.5cm]{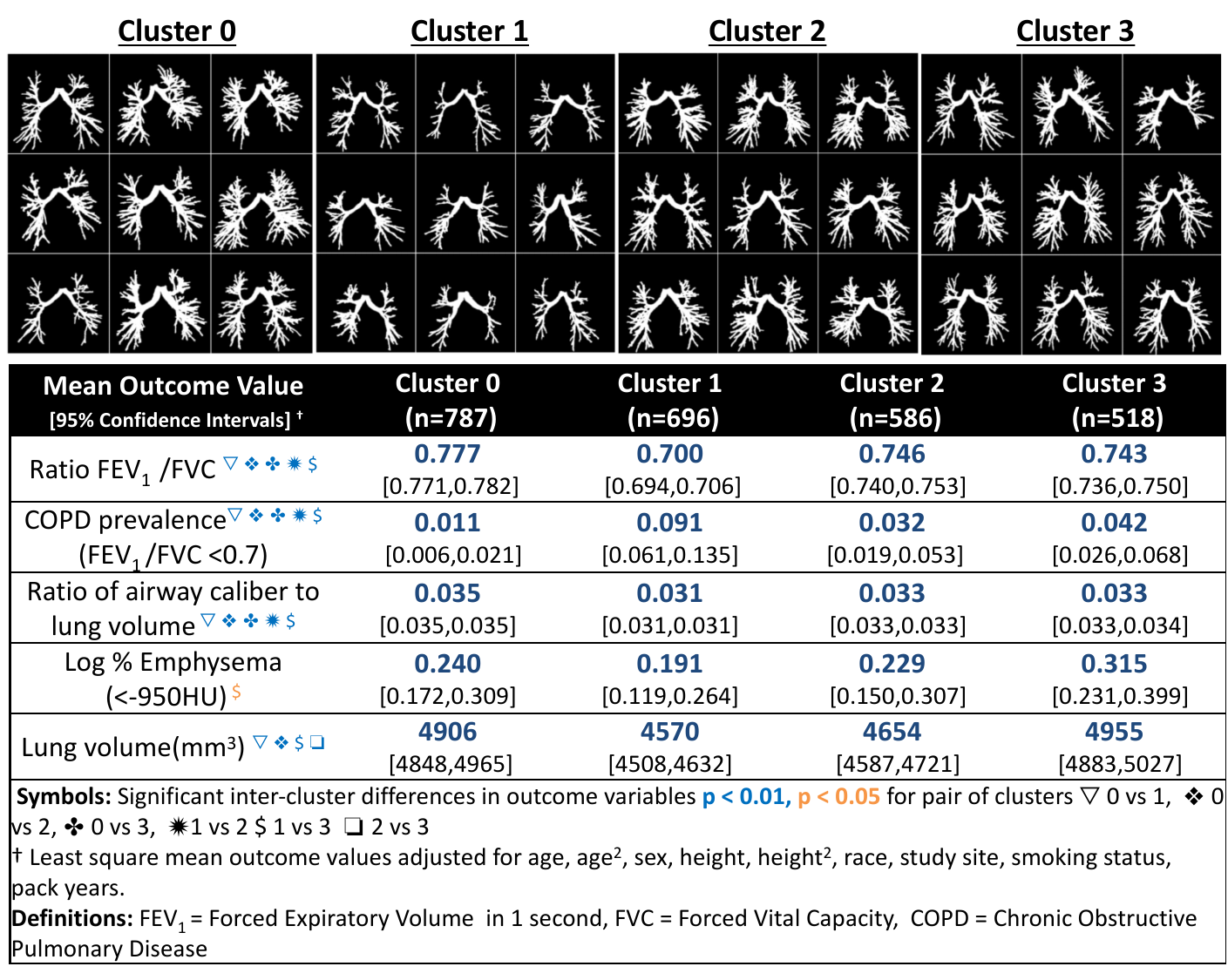}}}

  \centerline{(b) Random samples per cluster from $\mathcal{C}(D+NT)$}\medskip
\end{minipage}}

\caption{Per-cluster visualisation of non-dilated airway MIPs, with average cluster properties $(\mu [95\% CI])$. Significant inter-cluster differences in outcome variables are indicated (p-values \textcolor{blue_p}{$<0.01$ in blue}, \textcolor{orange_p}{$<0.05$ in orange}).}

\label{fig:cluster_contents}
\end{figure*}

\label{sec:pagestyle}
\subsection{Evaluation metrics}\label{subsec:eval_metrics}

\textbf{MIP Auto-encoding}: We compare the ground truth MIPS $\mathcal{M}$, and corresponding centerline skeleton $\mathcal{S}$ with their  versions ($\hat{\mathcal{M}},\hat{\mathcal{S}}$) from UNet-noskips encoding. We used the following metrics, as in \cite{garcia2021automatic}: Tree Length (TL) $= \frac{|\mathcal{S}\cap \hat{\mathcal{M}}|}{|\mathcal{S}|}$; Centerline leakage (CL)$=\frac{|\hat{\mathcal{S}}\not\in \mathcal{M}|}{|\mathcal{S}|}$; False positive rate (FPR) $=\frac{|\hat{\mathcal{M}}\not\in \mathcal{M}|}{|\mathcal{M}|}$; Dice score coefficient (DSC) = $\frac{2|\hat{\mathcal{M}} \cap \mathcal{M}|}{|\mathcal{M}| + |\hat{\mathcal{M}}|}$. We report mean and standard deviation of metrics across validation folds. We report average metrics across the three MIP views.

\textbf{MIP Clustering}: Agreement between two clusterings $(\mathcal{C}, \hat{\mathcal{C}})$ is evaluated using the Rand Index (RI) and Adjusted Rand Index (ARI). The Rand Index (RI) considers the number of pairs of datapoints that are co-clustered in both $\mathcal{C} \text{ and } \hat{\mathcal{C}}$ (denoted true positive (TP) samples), and pairs that belong to different clusters in both $\mathcal{C}$ and $\hat{\mathcal{C}}$ (denoted true negative (TN) samples) and is defined  $RI =\frac{|TP| + |TN|}{n(n+1)/2} $ where the denominator is the total number of pairs for $n$ data points. RI increases as the number of clusters increases.

ARI tackles this by adjusting RI such that ARI = 0 for uniform random assignment of data points across clusters and 1 for perfect agreement, $ARI = \frac{RI - \mathbb{E}(RI)}{1-\mathbb{E}(RI)}$.

\subsection{MIP Auto-encoding}\label{subsec:aw_results}
We present the mean/standard deviation values of MIP reconstruction metrics in MESA Exam 5 Table \ref{tab:results_pretrain} for different combinations of trachea inclusion and use of small airway dilation. 
When training with the trachea (T), all Dice scores are high. 
But, tree length remains low (TL $<70\%$) when validating on ND+T MIPs, emphasising the challenge of reconstructing fine airway structures at subject level, even if training on ND data.

Fine-tuning the best model (D+T/D+T) on dilated MIPs with the trachea masked out maintains performance in TL and FPR. Dice score falls marginally, likely due to the removal of the large, easy to encode trachea.

Evaluation on ND+T MIPs from the ATM'22 cohort demonstrates the robustness of our pre-trained autoencoder. All metrics are in line with MESA Exam 5, with Dice score outperforming at 0.835 in part due to smaller segmented trees.

\begin{table}
    \caption{MIP auto-encoding with UNet-noskips. We threshold output probability maps at 0.5 to generate binary masks. *Datatypes: D = MIPs with  peripheral airway dilation, ND = MIPS with no dilation, T = MIPS including trachea, NT = MIPS with masked trachea. Best model is indicated with $^{\dagger}$ and evaluated on the non-dilated ATM cohort (D+T $^{\dagger}$ / ND+T). We fine-tuned the best model using D+NT data for 5 epochs (FT$^{\dagger}$). Blue=evaluation on dilated images.
    }
    \centering
   \begin{center}
        \resizebox{0.99\linewidth}{!}{%
            \begin{tabular}{c|ccccccc}
                \toprule
  
                \multirow{2}{2.5cm}{Train/Eval Data} &\multicolumn{4}{c}{5 fold cross-validation ($\mu \pm \sigma$)} \\
                \cline{2-5}
                &Dice ($\uparrow$) &FPR ($\downarrow$) &TL ($\uparrow$) &CL ($\downarrow$) \\
                \hline
                          \addlinespace
                  \multicolumn{5}{c}{\textbf{MIPS}$\rightarrow$ \textbf{MIPS pre-training on MESA Exam 5}}\\	
                \hline
                 \rowcolor{babyblueeyes}
                D+T/D+T $^{\dagger}$&0.895±0.002
&0.023±0.001

&0.933±0.003

&0.011±0.000

 \\
                D+T/ND+T &0.793±0.006
&0.027±0.002

&0.6815±0.02
&0.021±0.001

 \\
 \rowcolor{babyblueeyes}
 ND+T/D+T &0.887±0.01
&0.034±0.003
&0.692±0.016
&0.021±0.002
\\
                ND+T/ND+T &0.798±0.01
&0.029±0.003
&0.692±0.016
&0.022±0.002
\\
\bottomrule
\addlinespace
         \multicolumn{5}{c}{\textbf{Fine-tuning best model$^{\dagger}$ on MESA Exam 5 (masked trachea)}}\\	
                \hline
                 \rowcolor{babyblueeyes}
                D+NT/D+NT (FT$^{\dagger}$) 
                &0.884±0.001
                &0.024±0.001
                &0.929±0.002
                &0.012±0.001 \\
                \hline
	\addlinespace
   \multicolumn{5}{c}{\textbf{Best model} $^{\dagger}$ \textbf{evaluation on ATM}
   }\\
      
 \bottomrule
                
              D+T $^{\dagger}$ /  ND+T 
                &0.835±0.034
                &0.029±0.007
                &0.773±0.07
                &0.02±0.008 \\

                \bottomrule
        \end{tabular}}
    \end{center}
    \label{tab:results_pretrain}
\end{table}

\subsection{Airway tree clustering}\label{subsec:clustering}

\begin{figure}[htb]
 \resizebox{\linewidth}{!}{

\begin{minipage}[b]{0.48\linewidth}
  \centering
\centerline{\includegraphics[width=4cm]{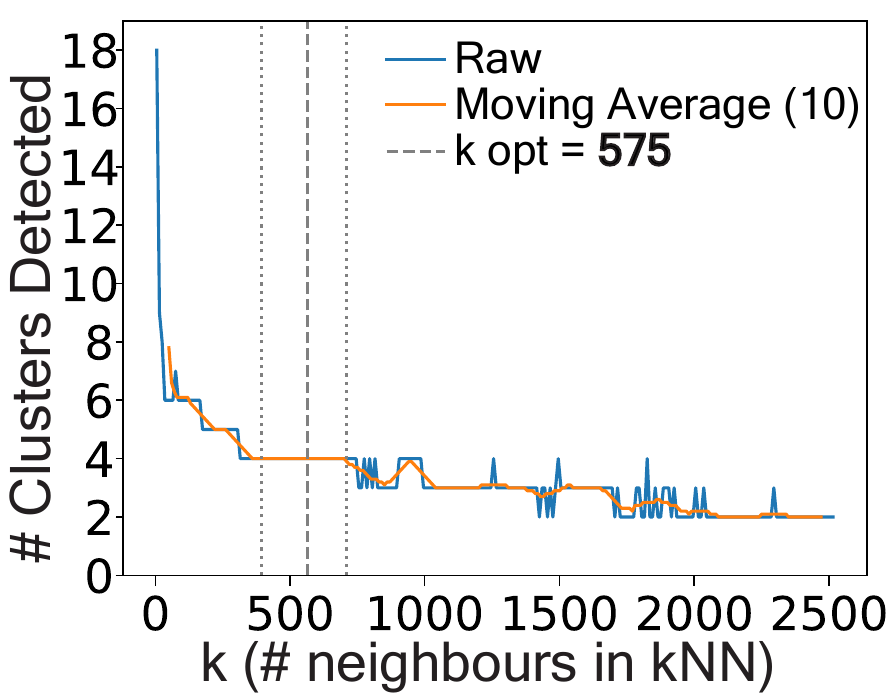}}

  \centerline{(a)}
  \medskip
\end{minipage}
\begin{minipage}[b]{0.48\linewidth}
  \centering
\centerline{\includegraphics[width=4.2cm]{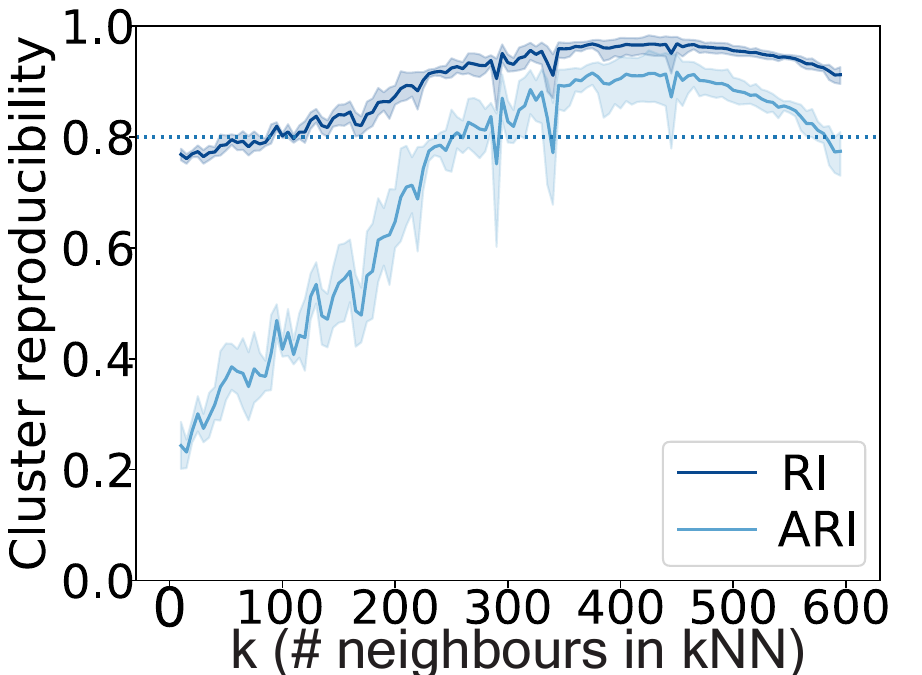}}

  \centerline{(b)}
  \medskip
\end{minipage}}

\caption{(a) Tuning $k_{opt}$=number of nearest neighbors used for clustering for  $\mathcal{C}(D+T)$. (b) Reproducibility of clusters using $k_{opt}=535$ versus $k$ across 5 subsets for $\mathcal{C}(D+NT)$. 
}

\label{fig:tune_kopt_reprod}

\end{figure}

\textbf{With trachea $\mathcal{C}(D+T)$: }
Randomly selected cases from the 4 discovered clusters (W,X,Y,Z) are shown in Figure \ref{fig:cluster_contents}a. Clusters seem driven by global shape features such as tree size and orientation of the trachea.
Repeated clustering on 5 subsets of MESA Exam 5 (80\% each), lead to $k_{opt} = 575$ (Figure \ref{fig:tune_kopt_reprod}a) and 4 clusters for each run. Comparing to clustering on 100\% of the data we got RI = \{0.956, 0.953, 0.944, 0.864, 0.974\} and ARI = \{0.851, 0.847, 0.834, 0.673, 0.901\}. Hence, cluster reproducibility is strong, despite standard deviation of the ARI being quite high. Comparing L2 distance against Cosine distance in the $k$NN graph (on 100\% of the training data) yields RI = 0.960, ARI = 0.865 indicating  robustness to $k$NN graph distance metric choice. 
Clustering with 1,024-D vs the 2,048-D feature vectors yields RI = 0.977, ARI = 0.942.
\textbf{Without trachea $\mathcal{C}(D+NT)$}: 
Clusters (0,1,2,3), illustrated in Figure \ref{fig:cluster_contents}b, show less distinct obvious global difference, indicating more subtle variations in airway trees being picked by the auto-encoder feature extractor.
Repeated clustering lead to $k_{opt} = 535$, and 4 clusters, with higher reproduciblity metrics than with the trachea: RI =\{0.949, 0.934, 0.951, 0.956, 0.944\},  ARI = \{0.869, 0.828, 0.873, 0.886, 0.855\}. Clustering output across the 5 subsets was stable across a wide range of $k$, indicating strong robustness to $k$NN graph construction (Figure \ref{fig:tune_kopt_reprod}b). 

We present the variation in clinical outcomes for  $\mathcal{C}(D+NT)$ in Figure \ref{fig:cluster_contents}b.
Cluster properties demonstrate the capabilities of deep MIP encoding to identify stable partitions with significant clinical differences (COPD prevalence in Cluster 0 is 0.011, while Cluster 1 is 8x higher) in the general population of MESA Exam 5. Cluster 1 has high COPD prevalence without high percent emphysema, which is an open area of clinical investigation and demonstrates power of using the airway segmentation rather than the CT intensities as input.

\section{Discussion}
\label{sec:discussion}
In this work, we have introduced and validated a framework for deep-learned representations of airway trees segmented on HRCT scans. We demonstrate the efficacy of dilating peripheral airways to improve depth of reconstructions achievable without the use of patch-based methods. We demonstrate for the first time, the use of unstructured, deep-learned shape features for the robust unsupervised clustering of airway trees to discover new phenotypes. We demonstrate that removal of the trachea results in more reproducible clusters, and reduces the impact of initial tree orientation on cluster assignment (without complex registration). Training on dilated airways was risky, but from our results, we see it is sufficient to discover sub-types with significant clinical associations with clinical COPD prevalence. This could drive future efforts in airway segmentation challenges to focus on tree depth rather than voxel level radius estimation. Our community discovery clustering method identifies four subtypes within MESA Exam 5 that are robust to the user-defined hyperparameters in our clustering pipeline and exhibit significant differences in COPD risk.\\
Although our deep-learned shape features indicate a strong understanding of airway tree structures, separation of specific topological variations are not fully identified due to the use of 2D MIPs representation with lower resolution and dilation of peripheral airways. Future work will explore multi-scale models to dig deeper into the peripheral airway tree and investigate the clustering of individual lobes of the lung. Our UNet-noskips model performs better on shallow trees, and outperforms in the coronal view. Future work will focus on hard-example mining to alleviate this imbalance. We further plan to validate our method in independent CT cohorts, e.g., Canadian Cohort Obstructive Lung Disease (CanCOLD)\cite{bourbeau2014canadian}, SubPopoulations and Intermediate Outcomes in COPD (SPIROMICS) study \cite{couper2014design}, and conduct longitudinal analysis in MESA Exams 6, 7.

\vfill
\pagebreak

\section{Compliance with ethical standards}
\label{sec:ethics}
Each MESA study site was approved by the institutional review board (http://www.mesa-nhlbi.org). Written informed consent was obtained from all participants. 

\section{Acknowledgments}
\label{sec:acknowledgments}

This research was supported by National Heart, Lung, and Blood Institute grants NIH R01-HL130506, R01 HL155816 \\NIH R01 HL121270, R01-HL077162 and R01-HL093081 and contracts 75N92020D00001,
HHSN268201500003I, \\ N01-HC-95159, 75N92020D00005,
N01-HC-95160, \\75N92020D00002, N01-HC-95161,
75N92020D00003, N01-HC-95162, 75N92020D00006,
N01-HC-95163, \\75N92020D00004, N01-HC-95164,
75N92020D00007,\\ N01-HC-95165, N01-HC-95166, N01-
HC-95167, \\N01-HC-95168 and N01-HC-95169, and by grants\\
UL1-TR-000040, UL1-TR-001079, and UL1-TR-001420
from the National Center for Advancing Translational
Sciences (NCATS). The authors thank the other investigators, the staff, and the participants of the MESA study for their valuable contributions. A full list of
participating MESA investigators and institutions can be found at http://www.mesa-nhlbi.org.
Author E.A.H is a shareholder of VIDA Diagnostics, Inc.

\bibliographystyle{IEEEbib}

\bibliography{strings}

\begin{thebibliography}{10}

\bibitem{vameghestahbanati2023association}
M.~Vameghestahbanati et~al.,
\newblock ``Association of dysanapsis with mortality among older adults,''
\newblock {\em European Respiratory Journal}, vol. 61, no. 6, 2023.

\bibitem{smith2018human}
B.~M. Smith et~al.,
\newblock ``Human airway branch variation and chronic obstructive pulmonary disease,''
\newblock {\em Proceedings of the National Academy of Sciences}, vol. 115, no. 5, pp. E974--E981, 2018.

\bibitem{smith2020association}
B.~M. Smith et~al.,
\newblock ``Association of dysanapsis with chronic obstructive pulmonary disease among older adults,''
\newblock {\em Journal of the American Medical Association}, vol. 323, no. 22, pp. 2268--2280, 2020.

\bibitem{bodduluri2021computed}
S.~Bodduluri et~al.,
\newblock ``Computed tomography--based airway surface area--to-volume ratio for phenotyping airway remodeling in chronic obstructive pulmonary disease,''
\newblock {\em American Journal of Respiratory and Critical Care Medicine}, vol. 203, no. 2, pp. 185--191, 2021.

\bibitem{bodduluri2018airway}
S.~Bodduluri et~al.,
\newblock ``Airway fractal dimension predicts respiratory morbidity and mortality in copd,''
\newblock {\em The Journal of Clinical Investigation}, vol. 128, no. 12, pp. 5374--5382, 2018.

\bibitem{lambert2011regional}
A.~R. Lambert et~al.,
\newblock ``Regional deposition of particles in an image-based airway model: large-eddy simulation and left-right lung ventilation asymmetry,''
\newblock {\em Aerosol Science and Technology}, vol. 45, no. 1, pp. 11--25, 2011.

\bibitem{billera2001geometry}
L.~J. Billera, S.~P. Holmes, and K.~Vogtmann,
\newblock ``Geometry of the space of phylogenetic trees,''
\newblock {\em Advances in Applied Mathematics}, vol. 27, no. 4, pp. 733--767, 2001.

\bibitem{wysoczanski2021unsupervised}
A.~Wysoczanski et~al.,
\newblock ``Unsupervised clustering of airway tree structures on high-resolution ct: the mesa lung study,''
\newblock in {\em Proceedings of International Symposium on Biomedical Imaging}, 2021, pp. 1568--1572.

\bibitem{feragen2013tree}
A.~Feragen et~al.,
\newblock ``Tree-space statistics and approximations for large-scale analysis of anatomical trees,''
\newblock in {\em Proceedings of Information Processing in Medical Imaging}, 2013, pp. 74--85.

\bibitem{nadeem2020ct}
S.~A. Nadeem et~al.,
\newblock ``A ct -based automated algorithm for airway segmentation using freeze-and-grow propagation and deep learning,''
\newblock {\em IEEE Transactions on Medical Imaging}, vol. 40, no. 1, pp. 405--418, 2020.

\bibitem{zhang2023multi}
M.~Zhang et~al.,
\newblock ``Multi-site, multi-domain airway tree modeling,''
\newblock {\em Medical Image Analysis}, vol. 90, pp. 102957, 2023.

\bibitem{bild2002multi}
D.~E. Bild et~al.,
\newblock ``Multi-ethnic study of atherosclerosis: objectives and design,''
\newblock {\em American journal of epidemiology}, vol. 156, no. 9, pp. 871--881, 2002.

\bibitem{rodriguez2010association}
J.~Rodriguez et~al.,
\newblock ``The association of pipe and cigar use with cotinine levels, lung function, and airflow obstruction: a cross-sectional study,''
\newblock {\em Annals of internal medicine}, vol. 203, no. 2, pp. 185--191, 2010.

\bibitem{sieren2016spiromics}
J.~P. Sieren et~al.,
\newblock ``{SPIROMICS} protocol for multicenter quantitative computed tomography to phenotype the lungs,''
\newblock {\em American journal of respiratory and critical care medicine}, vol. 194, no. 7, pp. 794--806, 2016.

\bibitem{ronneberger2015u}
O.~Ronneberger, Philipp Fischer, and T.~Brox,
\newblock ``U-net: convolutional networks for biomedical image segmentation,''
\newblock in {\em Proceedings of Medical Image Computing and Computer-Assisted Intervention}, 2015, pp. 234--241.

\bibitem{de2011generalized}
P.~De~Meo et~al.,
\newblock ``Generalized louvain method for community detection in large networks,''
\newblock in {\em International conference on intelligent systems design and applications}, 2011, pp. 88--93.

\bibitem{garcia2021automatic}
A.~Garcia-Uceda et~al.,
\newblock ``Automatic airway segmentation from computed tomography using robust and efficient 3-d convolutional neural networks,''
\newblock {\em Scientific Reports}, vol. 11, no. 1, pp. 16001, 2021.

\bibitem{bourbeau2014canadian}
J.~Bourbeau et~al.,
\newblock ``Canadian cohort obstructive lung disease ({CanCOLD}): fulfilling the need for longitudinal observational studies in copd,''
\newblock {\em COPD: Journal of Chronic Obstructive Pulmonary Disease}, vol. 11, no. 2, pp. 125--132, 2014.

\bibitem{couper2014design}
D.~Couper et~al.,
\newblock ``Design of the subpopulations and intermediate outcomes in {COPD} study ({SPIROMICS}),''
\newblock {\em Thorax}, vol. 69, no. 5, pp. 492--495, 2014.

\end{thebibliography}

\end{document}